\documentclass[a4paper]{article}

\usepackage{INTERSPEECH2019}
\usepackage[colorlinks,citecolor=red,urlcolor=blue,bookmarks=false,hypertexnames=true]{hyperref}
\usepackage{cleveref}

\usepackage{esvect}
\usepackage{booktabs,caption}
\usepackage[flushleft]{threeparttable}

\usepackage[shortlabels]{enumitem}

\newcommand{\bbmm}[1]{#1}

\title{A Convolutional Deep Markov Model\\for Unsupervised Speech Representation Learning}
\name{\small Sameer Khurana$^1$, Antoine Laurent$^2$, Wei-Ning Hsu$^1$, Jan Chorowski$^3$, Adrian Lancucki$^4$, Ricard Marxer$^5$, James Glass$^1$}
\address{ \small
  $^1$MIT - CSAIL, Cambridge, USA\\  $^2$LIUM - Le Mans University, France\\ $^3$University of Wrocław, Poland\quad 
  $^4$NVIDIA Corporation, Poland \quad $^5$LIS - University of Toulon, France }
\email{skhurana@mit.edu, antoine.laurent@univ-lemans.fr}

\begin{document}

\maketitle
\begin{abstract}
Probabilistic Latent Variable Models (LVMs) provide an alternative to self-supervised learning approaches for linguistic representation learning from speech. LVMs admit an intuitive probabilistic interpretation where the latent structure shapes the information extracted from the signal. Even though LVMs have recently seen a renewed interest due to the introduction of Variational Autoencoders (VAEs), their use for speech representation learning remains largely unexplored. In this work, we propose  Convolutional Deep Markov Model (ConvDMM), a Gaussian state-space model with non-linear emission and transition functions modelled by deep neural networks. This unsupervised model is trained using black box variational inference. A deep convolutional neural network is used as an inference network for structured variational approximation. When trained on a large scale speech dataset (LibriSpeech), ConvDMM produces features that significantly outperform multiple self-supervised feature extracting methods on linear phone classification and recognition on the Wall Street Journal dataset. Furthermore, we found that ConvDMM complements self-supervised methods like Wav2Vec and PASE, improving on the results achieved with any of the methods alone. Lastly, we find that ConvDMM features enable learning better phone recognizers than any other features in an extreme low-resource regime with few labelled training examples.
\end{abstract}

\section{Introduction}



One of the long-standing goals of speech and cognitive scientists is to develop a computational model of language acquisition \cite{goldwater2009bayesian, lee2012nonparametric, ondel2016variational, harwath2016unsupervised}. Early on in their lives, human infants learn to recognize phonemic contrasts, frequent words and other linguistic phenomena underlying the language \cite{dupoux2018cognitive}. The computational modeling framework of generative models is well-suited for the problem of spoken language acquisition, as it relates to the classic analysis-by-synthesis theories of speech recognition \cite{halle1962speech, liberman1967perception}. Although, generative models are theoretically elegant and informed by theories of cognition, most recent success in speech representation learning has come from self-supervised learning algorithms such as Wav2Vec \cite{schneider2019wav2vec}, Problem Agnostic Speech Encoding (PASE) \cite{pascual2019learning}, Autoregressive Predictive Coding (APC) \cite{chung2019unsupervised}, MockingJay (MJ) \cite{liu2019mockingjay} and Deep Audio Visual Embedding Network (DAVENet) \cite{harwath2020learning}. Generative models present many advantages with respect to their discriminative counterparts. They have been used for disentangled representation learning in speech \cite{hsu2017unsupervised, khurana2019factorial, li2018disentangled}. Due to the probabilistic nature of these models, they can be used for generating new data and hence, used for data augmentation \cite{hsu2017unsuperviseda, hsu2018unsupervised} for Automatic Speech Recognition (ASR), and anomaly detection \cite{Grathwohl2020Your}.

In this paper, we focus solely on designing a generative model for low-level linguistic representation learning from speech.
We propose Convolutional Deep Markov Model (ConvDMM), a Gaussian state-space model with non-linear emission and transition functions parametrized by deep neural networks and a Deep Convolutional inference network. The model is trained using amortized black box variational inference (BBVI) \cite{ranganath2013black}. Our model is directly based on the Deep Markov Model proposed by Krishnan et. al \cite{krishnan2017structured}, and draws from their general mathematical formulation for BBVI in non-linear Gaussian state-space models. When trained on a large speech dataset, ConvDMM produces features that outperform multiple self-supervised learning algorithms on downstream phone classification and recognition tasks, thus providing a viable latent variable model for extracting linguistic information from speech.

We make the following contributions:
\begin{enumerate}[1)]
    \item Design a generative model capable of learning good quality linguistic representations, which is competitive with recently proposed self-supervised learning algorithms on downstream linear phone classification and recognition tasks.
    \item Show that the ConvDMM features can significantly outperform other representations in linear phone recognition, when there is little labelled speech data available.
    \item Lastly, demonstrate that by modeling the temporal structure in the latent space, our model learns better representations compared to assuming independence among latent states.
\end{enumerate}

\section{The Convolutional Deep Markov Model}
\label{sec:2}
\subsection{ConvDMM Generative Process} \label{sec:2.1}  Given the functions; $\bbmm{f_{\theta}(\cdot), u_{\theta}(\cdot)}$ and $\bbmm{t_{\theta}(\cdot)}$, the ConvDMM generates the sequence of observed random variables, $\bbmm{x}_{1:T} = \bbmm{(x}_1, \ldots,\bbmm{x}_T)$, using the following generative process
\begin{align}
    \bbmm{z}_1 &\sim \bbmm{\mathcal{N}(0, I)}\\
    \bbmm{z}_\tau|\bbmm{z}_{\tau-1} &\sim \bbmm{\mathcal{N}(t^{\mu}_{\theta}(z}_{\tau-1}),\bbmm{ t^{\sigma}_{\theta}(z}_{\tau-1}))& \tau=2,\ldots,L \label{eq:2}\\
    \bbmm{e}_{1:T} &= \bbmm{u_\theta(z}_{1:L}) \\
    \bbmm{\mu}_{1:T} &= \bbmm{f_{\theta}(e}_{1:T})\\
    \bbmm{x}_t|\bbmm{e}_t &\sim \bbmm{\mathcal{N}(\mu}_t, \gamma)& t=1,\ldots, T \label{eq:5}
\end{align}
where $T$ is a multiple of $L$, $T = k\cdot L$ and $\bbmm{z}_{1:L}$ is the sequence of latent states. We assume that the observed and latent random variables come from a multivariate normal distribution with diagonal covariances. 
The joint density of latent and observed variables for a single sequence is
\begin{equation}
    \bbmm{p(z}_{1:L}, \bbmm{x}_{1:T}) = \bbmm{p(x}_{1:T}|\bbmm{z}_{1:L}) \bbmm{p(z}_1)\prod\limits_{\tau=2}^{L} \bbmm{p(z}_\tau|\bbmm{z}_{\tau-1}).
\end{equation}
For a dataset of i.i.d. speech sequences, the total joint density is simply the product of per sequence joint densities.
The scale $\bbmm{\gamma}$ is learned during training.

\textbf{The transition function} $\bbmm{t_\theta : z}_{\tau-1} \rightarrow \bbmm{(\mu}_\tau, \bbmm{\sigma}_\tau) \big|_{\tau=2}^{L}$ estimates the mean and scale of the Gaussian density over the latent states. It is implemented as a Gated Feed-Forward Neural Network~\cite{krishnan2017structured}.
The gated transition function could capture both linear and non-linear transitions. 

\textbf{The embedding function} $\bbmm{u_{\theta} : z}_{1:L} \rightarrow \bbmm{e}_{1:T}$ transforms and up-samples the latent sequence to the same length as the observed sequence. 
It is parametrized by a four layer CNN with kernels of size 3, 1024 channels and residual connections. We use the activations of the last layer of the embedding CNN as the features for the downstream task. This is reminiscent of kernel methods \cite{hofmann2008kernel} where the raw input data are mapped to a high dimensional feature space using a user specified feature map. In our case, the CNN plays a similar role, mapping the low-dimensional latent vector sequence, $\bbmm{z}_{1:L} \in \mathbb{R}^{L \times 16}$, to a high dimensional vector sequence, $\bbmm{e}_{1:T}\in  \mathbb{R}^{T\times 1024}$, by repeating the output activations of the CNN $k$ times, where $k = T/L$. In our case, $k$ is 4 which is also the downsampling factor of the encoder function (\S~\ref{sec:inf}). A similar module was used in Chorowski et. al \cite{chorowski2019unsupervised}, where they used a single CNN layer after the latent sequence.

\textbf{The emission function} $\bbmm{f_{\theta} : e}_t \rightarrow \bbmm{(\mu}_t)\big|_{t=1}^{T}$ (a decoder) estimates the mean of the likelihood function.
It is a two-layered MLP with 256 hidden units and residual connections. We employ a low capacity decoder to avoid the problem of posterior collapse \cite{bowman2015generating}, a common problem with high capacity decoders. 
 
\subsection{ConvDMM Inference} \label{sec:inf} The goal of inference is to estimate the posterior density of latent random variables given the observations $\bbmm{p(z | x)}$. Exact posterior inference in non-conjugate models like ConvDMM is intractable, hence we turn to Variational Inference (VI) for approximate inference. We use  VI and BBVI interchangeably throughout the rest of the paper. In VI, we approximate the intractable posterior $\bbmm{p(z|x)}$ with a tractable family of distributions, known as the variational family $\bbmm{q_{\phi}(z|x)}$, indexed by $\phi$. In our case, the variational family takes the form of a Gaussian with diagonal covariance. Next, we briefly explain the Variational Inference process for ConvDMM.

Given a realization of the observed random variable sequence $\bbmm{x}_{1:T} = \bbmm{x}_1, \ldots, \bbmm{x}_T$, the initial state parameter vector $\bbmm{\hat{z}_1}$, and the functions $\bbmm{g_{\phi}(\cdot)}$ and $\bbmm{c_{\phi}(\cdot)}$, the process of estimating the latent states can be summarized as:
\begin{align}
    \bbmm{h}_{1:L} &= \bbmm{g_{\phi}(x}_{1:T}) \\
    \bbmm{\hat{z}}_\tau|\bbmm{\hat{z}}_{\tau-1}, \bbmm{x} &\sim \bbmm{\mathcal{N}(c^{\mu}_{\phi}(h}_\tau, \bbmm{\hat{z}}_{\tau-1}),\bbmm{ c^{\sigma}_{\phi}(h}_\tau,\bbmm{ \hat{z}}_{\tau-1}))& \tau=2\ \text{to}\ L .
\end{align}
Let $T = k*L$, where $k$ is the down-sampling factor of the encoder, $\bbmm{g_{\phi} : x}_{1:T} \rightarrow \bbmm{h}_{1:L}$ is the encoder function, $\bbmm{c_{\phi}}$ is the combiner function that provides posterior estimates for the latent random variables. 

We parameterize the encoder $\bbmm{g_\phi}$ using a 13-layer CNN with kernel sizes (3, 3, 3, 3, 3, 4, 4, 3, 3, 3, 3, 3, 3), strides (1, 1, 1, 1, 1, 2, 2, 1, 1, 1, 1, 1, 1) and 1024 hidden channels. The encoder down-samples the input sequence by a factor of four. The last layer of the encoder with $\bbmm{h}_{1:L}$ as its hidden activations has a receptive field of approximately 50. This convolutional architecture is inspired by \cite{chorowski2019unsupervised}, but other acoustic models such as Time-Depth Separable Convolutions~\cite{hannun2019sequence}, VGG transformer~\cite{mohamed2019transformers}, or ResDAVENet~\cite{harwath2020learning} could be used here. We leave this investigation for future work.

\textbf{The combiner function} $\bbmm{c_\phi}$ provides structured approximations of the variational posterior over the latent variables by taking into account the prior Markov latent structure. The combiner function follows~\cite{krishnan2017structured}:
\begin{align*}
    \bbmm{h_{\text{combined}}} &= \bbmm{\frac{1}{2} (\text{tanh}(W\hat{z}}_{t-1} + \bbmm{b) + h}_t)\\
    \bbmm{\mu}_t &= \bbmm{W_{\mu} h_{\text{combined}} + b_{\mu}}\\
    \bbmm{\sigma}_t &= \bbmm{\text{softplus}(W_{\sigma}h_{\text{combined}} + b_{\sigma})}.
\end{align*}
It uses tanh non-linearity on $\bbmm{z_{t-1}}$ to approximate the transition function. Future work could investigate sharing parameters with the generative model as in Maal{\o}e et. al's Bidirection inference VAE (BIVA) \cite{maaloe2019biva}. We note that structured variational inference in neural variational models is an important area of research in machine learning, with significant recent developemnts~\cite{johnson2016composing, lin2018variational}. Structured VAE has also been used for acoustic unit discovery~\cite{ebbers2017hidden}, which is not the focus of this work.

\begin{figure}
    \hspace{0.15cm}
    \includegraphics[width=\linewidth]{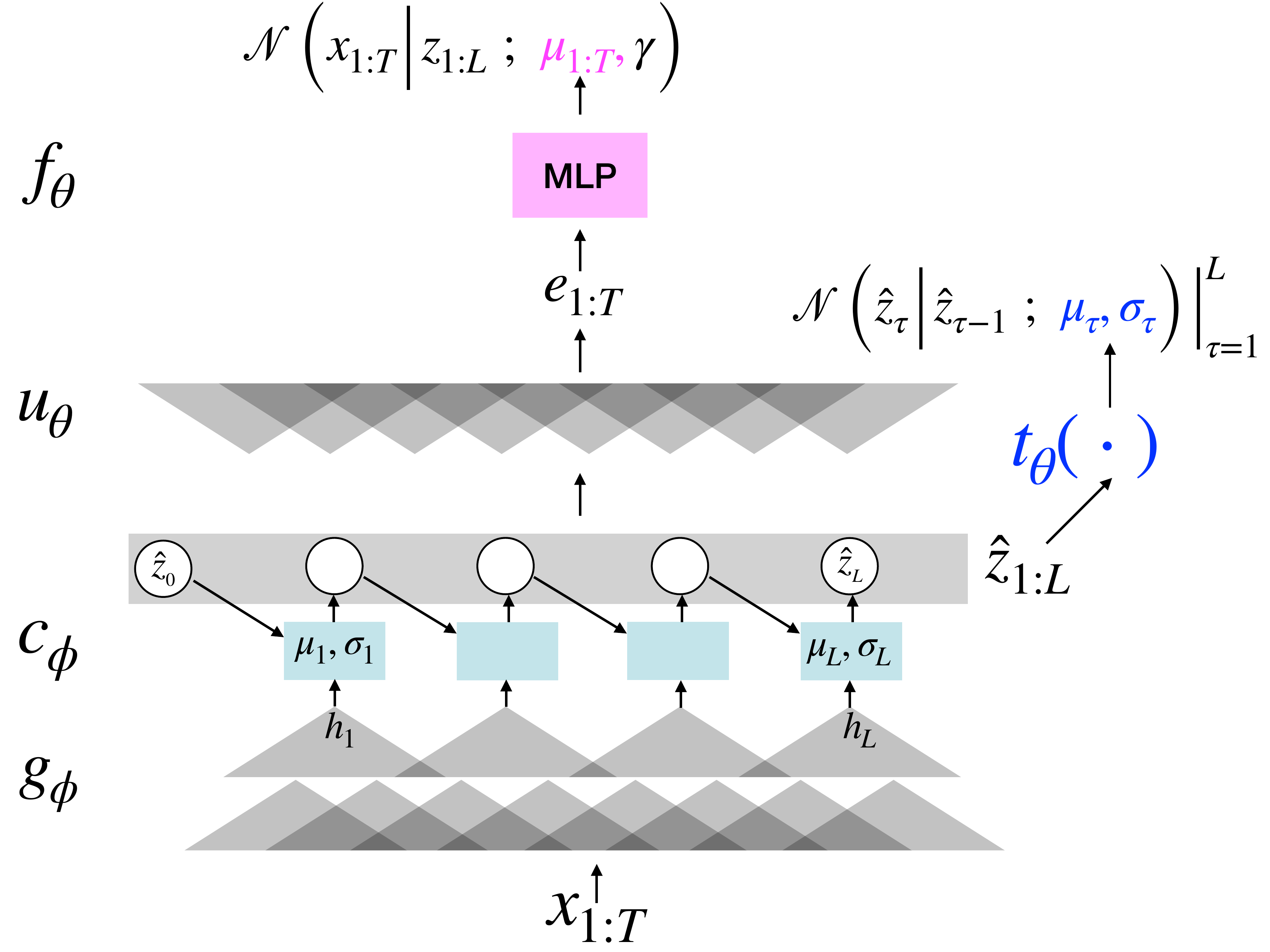}
    \caption{An illustration of the ConvDMM.}
    \label{fig:dmm}
\end{figure}

\begin{table*}[tb]
    \caption{Phone Classification (FER) and Recognition (PER) when trained on a subset of Wall Street Journal.
             Suffixes \emph{-50, -360, -960} denote the amount of hours of LibriSpeech training data used for training the unsupervised model.}
    \label{tab:wsj_probes}
    \centering
    \resizebox{\linewidth}{!}{%
    \begin{threeparttable}
    \begin{tabular}{llrrrrrrrrrrrrrrrrrr}\toprule
       \% of Labeled Data&&\multicolumn{2}{c}{1\%}&&\multicolumn{2}{c}{2\%}&&\multicolumn{2}{c}{5\%}&&\multicolumn{2}{c}{10\%}&&\multicolumn{2}{c}{50\%}&&\multicolumn{1}{c}{Low Shot (0.1\%)}\\
       \cmidrule{3-4} \cmidrule{6-7}\cmidrule{9-10}\cmidrule{12-13}\cmidrule{15-16}\cmidrule{18-18}
       &&FER&PER&&FER&PER&&FER&PER&&FER&PER&&FER&PER&&PER\\
       \midrule
       GaussVAE-960 &&-&55.8&&-&50.1&&-&48.2&&-&45.9&&-&42.5&&- \\
       Supervised Transfer-960 &&-&17.9&&-&16.4&&-&14.4&&-&12.8&&-&10.8&&25.8 ($\pm$ 0.96) \\
       \midrule
       \multicolumn{16}{l}{\textbf{\textit{Self Supervised Learning}:}}\\
       MockingJay-960 \cite{liu2019mockingjay} &&40.0&53.2&&38.5&48.8&&37.5&45.5&&37.0&44.2&&36.7&43.5&&-& \\
       PASE-50~\cite{pascual2019learning} &&34.7&61.2&&33.5&50.0&&33.2&49.0&&32.8&49.0&&32.7&48.2&&80.7 ($\pm$ 2.65)\\
       Wav2Vec-960 \cite{schneider2019wav2vec} &&19.8&37.6&&19.1&27.7&&18.8&24.5&&18.6&23.9&&18.5&22.9&&78.0 ($\pm$ 10.4)\\
       \midrule
       \multicolumn{16}{l}{\textbf{\textit{Audio-Visual Self Supervised Learning}:}}\\
       RDVQ (Conv2) \cite{harwath2020learning} &&31.6&44.1&&30.8&42.4&&30.5&41.1&&30.1&41.3&&30.2&40.6&&52.6 ($\pm$ 0.95)\\
       \midrule
       \multicolumn{16}{l}{\textbf{\textit{Proposed Latent Variable Model}:}}\\
       ConvDMM-50  &&29.6&37.8&&28.6&35.4&&27.9&31.3&&27.9&30.3&&27.0&29.1&&-\\
       ConvDMM-360  &&28.2&34.8&&27.0&30.8&&26.4&28.2&&25.9&27.7&&25.7&26.7&&-\\
       ConvDMM-960 &&27.7&32.5&&26.6&30.0&&26.0&28.1&&26.0&27.1&&25.6&26.0&&50.7 ($\pm$ 0.57) \\
       \midrule
       \multicolumn{16}{l}{\textbf{\textit{Modeling PASE and Wav2Vec features using the proposed generative model}:}}\\
       ConvDMM-960-PASE-50 &&-&35.4&&-&32.6&&-&30.6&&-&29.3&&-&28.4&&55.3 ($\pm$ 3.21)\\
       ConvDMM-Wav2Vec-960  &&-&28.6&&-&25.7&&-&22.3&&-&21.2&&-&20.4&&40.7 ($\pm$ 0.42)\\
       \midrule
    \end{tabular}
    \begin{tablenotes}
      \small
      \item[1] $\pm$ refers to the standard deviation in the results
    \end{tablenotes}
    \end{threeparttable}
    }
   
\end{table*}

\subsection{ConvDMM Training} ConvDMM like other VAEs is trained to maximize the bound on model likelihood, known as the Evidence Lower Bound (ELBO):
\begin{equation*}
	\bbmm{\mathcal{L(\theta, \phi)}} = \bbmm{E_{q_{\phi}(z|x)} [\text{log}\ p_{\theta}(x|z)] - \text{KL} (q_{\phi}(z|x) || p_{\theta}(z))}
\end{equation*}
where $\bbmm{p_\theta (x|z)} = \prod\limits_{t=1}^{T} \bbmm{p(x}_t|\bbmm{e}_t)$ is the Gaussian likelihood function and $\bbmm{p(x}_t|\bbmm{e}_t)$ is given by the ConvDMM generative process in Section~\ref{sec:2.1}.  The Gaussian assumption lets us use the \textit{reparametrization trick} \cite{kingma2013auto} to obtain low-variance unbiased Monte-Carlo estimates of the expected log-likelihood, the first term in the R.H.S of the ELBO. The KL term, which is also an expectation can be computed similarly and its gradients can be obtained analytically. In our case, we use the formulation of Equation 12, Appendix A., in Krishnan et. al \cite{krishnan2017structured}, to compute the KL term analytically.

The model is trained using the Adam optimizer with a learning rate of 0.001 for 100 epochs. We reduce the learning rate to half of its value if the loss on the development set plateaus for three consecutive epochs. L2 regularization on model parameters with weight 5e-7 is used during training. To avoid latent variable collapse we use KL annealing \cite{bowman2015generating} with a linear schedule, starting from an initial value of 0.5, for the first 20 epochs of training. We use a mini-batch size of 64 and train the model on a single NVIDIA Titan X Pascal GPU.

\section{Experiments}
\subsection{Evaluation Protocol and Dataset} We evaluate the learned representations on two tasks; phone classification and recognition. For phone classification, we use the ConvDMM features, the hidden activations from the last layer of the embedding function, as input to a softmax classifier, a linear projection followed by a softmax activation. The classifier is trained using Categorical Cross Entropy to predict framewise phone labels. For phone recognition the ConvDMM features are used as input to a softmax layer which is trained using Connectionist Temporal Classification (CTC) \cite{graves2006connectionist} to predict the output phone sequence. We do not fine-tune the ConvDMM feature extractor on the downstream tasks. The performance on the downstream tasks is driven solely by the learned representations as there is just a softmax classifier between the representations and the labels. The evaluation protocol is inspired by the unsupervised learning works in the Computer Vision community \cite{tian2019contrastive, henaff2019data}, where features extracted from representation learning systems trained on ImageNet are used as input to a softmax classifier for object recognition. Neural Networks for supervised learning have always been seen as feature extractors that project raw data into a linearly separable feature space making it easy to find decision boundaries using a linear classifier. We believe that it is reasonable to expect the same from unsupervised representation learning methods and hence, we compare all the representation learning methods using the aforementioned evaluation protocol. 

We train ConvDMM on the publicly available LibriSpeech dataset \cite{panayotov2015librispeech}. To be comparable with other representation learning methods with respect to the amount of training dataset used, we train another model on a small 50 hours subset of LibriSpeech. For evaluation, we use the Wall Street Journal (WSJ) dataset \cite{paul1992design}.

\begin{figure}
    \centering
    \includegraphics[width=\linewidth]{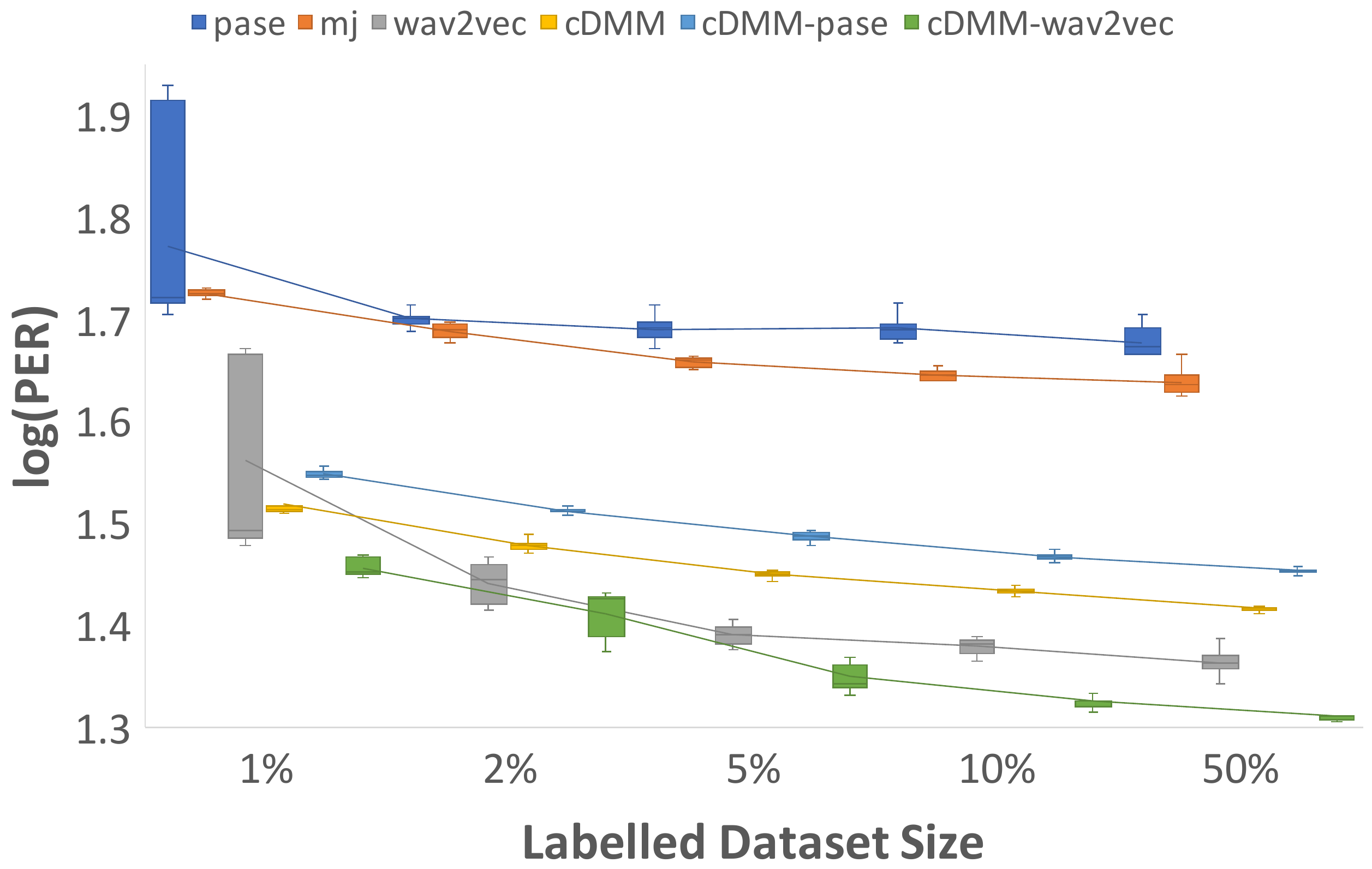}
    \caption{PER on WSJ eval92 dataset using features extracted from different models.}
    \label{box}
\end{figure}
\subsection{Results \& Discussion} Table~\ref{tab:wsj_probes} presents framewise linear phone classification (FER) and recognition (PER) error rates on the WSJ eval92 dataset for different representation learning techniques. ConvDMM is trained on Mel-Frequency Cepstral Coefficients (MFCCs) with concatenated delta and delta-delta features. ConvDMM-50 and PASE-50 are both trained on 50 hours of LibriSpeech, ConvDMM-960, Wav2Vec-960 and MockingJay-960 are trained on 960 hours. ConvDMM-360 is trained on the 360 hours of the clean LibriSpeech dataset. RDVQ is trained on the Places400k spoken caption dataset \cite{harwath2016unsupervised}.  We do not train any of the representation learning systems that are compared against ConvDMM on our own. We use the publicly available pre-trained checkpoints to extract features. The linear classifiers used to evaluate the features extracted from unsupervised learning systems are trained on different subsets of WSJ train dataset, ranging from 4 mins (0.1\%) to 40 hours (50\%).
To study the effect of modeling temporal structure in the latent space as in ConvDMM, we train a Gauss VAE which is similar to the ConvDMM except that it does not contain the transition model and hence, is a traditional VAE with  isotropic Gaussian priors over the latent states \cite{kingma2013auto}.

To generate the numbers in the table we perform the following steps.  Consider, for example, the column labelled 1\% as we describe how the numbers are generated for different models (rows). We randomly pick 1\% of the speech utterances in the WSJ train dataset. This is performed three times with different random seeds, yielding three different 1\% data splits of labelled utterances from the WSJ train dataset. We then train linear classifiers on the features extracted using different representation learning systems, on each of the three splits five times with different random seeds. This gives us a total of 15 classification and recognition error rates. The final number is the mean of these numbers after removing the outliers. Any number greater than $q_3 + 1.5*iqr$ or less than $q_1 - 1.5*iqr$, where $q_1$ is the first Quartile, $q_3$ is the third Quartile and $iqr$ is the inter-quartile range, is considered an outlier.  We follow the same procedure to create different training splits, 2\%, 5\%, 10\%, 50\%, from the WSJ train dataset and present classification error rates in the table for all splits. Figure~\ref{box} shows the box plot for the PER on WSJ eval92 dataset using features extracted from different models. 

In terms of PER, ConvDMM-50 outperforms PASE by 23.4 percentage points (pp), MockingJay by 15.4pp and RDVQ by 6.3pp under the scenario when 1\% of labeled training data is available to the linear phone recognizer, which corresponds to approximately 300 spoken utterances ($\approx$ 40 mins). Compared to Wav2Vec, ConvDMM lags by 0.2pp, but the variance in Wav2Vec results is very high as can be seen in Figure~\ref{box}. Under the 50\% labeled data scenario, ConvDMM-50 outperforms MockingJay by 14.4pp, PASE by 19.1pp, RDVQ by 11.5pp and lags Wav2Vec by 6.2pp. The gap between ConvDMM-50 and RDVQ widens in the 50\% labeled data case. ConvDMM-960 similarly outperforms all the methods under the 1\% labeled data scenario, outperforming Wav2Vec, the second best method, by 5.1pp. Also the variance in the ConvDMM-960 results is much lower than Wav2Vec (See Figure~\ref{box}). ConvDMM systematically outperforms the Gauss VAE which does not model the latent state transitions, showing the value of prior structure.

ConvDMM-PASE which is the ConvDMM model built on top of PASE features instead of the MFCC features, outperforms PASE features by 25.8pp under the 1\% labeled  data scenario. A significant gap exists under all data scenarios. Similar results can be observed with ConvDMM-Wav2Vec model, but the improvements over Wav2Vec features is not as drastic, probably due to the fact that Wav2Vec already produces very good features. For low shot phone recognition with 0.1\% labeled ($\approx$ 4 mins), ConvDMM-960 significantly outperforms all other methods. Surprisingly, RDVQ shows excellent performance under this scenario. ConvDMM-Wav2Vec-960 performs 10pp better than ConvDMM-960 trained on MFCC features and 38pp better than Wav2Vec features alone. We could not get below 90\% PER with MockingJay and hence, skip reporting the results.

Lastly, we compare the performance of features extracted using unsupervised learning systems trained on LibriSpeech vs features extracted using the fully supervised system neural network acoustic model trained on the task of phone recognition on 960 hours of labeled data (See the row labeled Supervised Transfer-960). The supervised system has the same CNN encoder as the ConvDMM. There is a glaring gap between the supervised system and all other representation learning techniques, even in the very few data regime (0.1\%). This shows there is still much work to be done in order to reduce this gap.

 
\section{Related Work}
\label{sec:5}

Another class of generative models that have been used to model speech but not explored in this work are the autoregressive models. Autoregressive models, a class of explicit density generative models, have been used to construct speech density estimators. Neural Autoregressive Density Estimator (NADE) \cite{uria2016neural} is a prominent earlier work followed by more recent Wavenet \cite{oord2016wavenet}, SampleRNN \cite{mehri2016samplernn} and MelNet \cite{vasquez2019melnet}. An interesting avenue of future research is to probe the internal representations of these models for linguistic information. We note that, Waveglow, a flow based generative model is recently proposed as an alternative to autoregressive models for speech \cite{prenger2019waveglow}.

\section{Conclusions}
In this work, we design the Convolutional Deep Markov Model (ConvDMM), a Gaussian state-space model with non-linear emission and transition functions parametrized by deep neural networks.
The main objective of this work is to demonstrate that generative models can reach the same, or even better, performance than self supervised models. 
In order to do so, we compared the ability of our model to learn linearly separable representations, by evaluating each model in terms of PER and FER using a simple linear classifier.
Results show that our generative model produces features that outperform multiple self-supervised learning methods on phone classification and recognition task on Wall Street Journal. We also find out that these features can achieve better performances than all other evaluated features when learning the phone recogniser with very few labelled training examples.
Another interesting outcome of this work is that by using self-supervised extracted features as input of our generative model, we produce features that outperforms every other one in the phone recogniser task. Probably due to enforcing temporal structure in the latent space.
Lastly, we argue that features learned using unsupervised methods are significantly worse than features learned by a fully supervised deep neural network acoustic model, setting the stage for future work.


\bibliographystyle{IEEEtran}

\bibliography{mybib}


\end{document}